\pdfoutput=1

\documentclass[11pt]{article}

\usepackage[preprint]{acl}

\usepackage{times}
\usepackage{latexsym}

\usepackage[T1]{fontenc}

\usepackage[utf8]{inputenc}

\usepackage{microtype}
\usepackage{amsmath}
\usepackage{multirow}

\usepackage{inconsolata}

\usepackage{graphicx}

\title{Improving Dense Passage Retrieval with Multiple Positive Passages}

\author{Shuai Chang \\
  China Academy of Railway Sciences Corporation Limited, Beijing \\
  \texttt{rocketduck@163.com} }

\begin{document}
\maketitle
\begin{abstract}
	
By leveraging a dual encoder architecture, Dense Passage Retrieval (DPR) has outperformed traditional sparse retrieval algorithms such as BM25 in terms of passage retrieval accuracy. Recently proposed methods have further enhanced DPR's performance. However, these models typically pair each question with only one positive passage during training, and the effect of associating multiple positive passages has not been examined. In this paper, we explore the performance of DPR when additional positive passages are incorporated during training. Experimental results show that equipping each question with multiple positive passages consistently improves retrieval accuracy, even when using a significantly smaller batch size, which enables training on a single GPU.

\end{abstract}

\section{Introduction}

In information retrieval (IR), passage retrieval refers to the task of retrieving text segments or passages that are relevant to a given query. Due to its ability to narrow down the searching scope, passage retrieval has become a key component in open-domain question answering (QA) and web search engines. Traditional methods such as TF-IDF and BM25 \citep{Robertson1997} rely on term frequency to measure text relevance, but lack the ability to capture the semantic meaning of sentences. This limitation could lead to poor performance when relative contents are composed of entirely different tokens \citep{Karpukhin2020}.

Pre-trained language models, such as BERT \citep{Devlin2019} and T5 \citep{Raffel2020}, have significantly enhanced text representation learning and demonstrated superior performance in IR tasks \citep{Nogueira2020, Ni2022}. The Dense Passage Retrieval (DPR) model \citep{Karpukhin2020} employs a dual-BERT encoder architecture to independently encode questions and passages into dense vector representations and utilizes in-batch negatives to improve training efficiency. DPR outperforms traditional sparse retrieval methods like BM25, thereby boosting the performance of end-to-end QA systems and retrieval-augmented generation models \citep{Lewis2020}. While several refinement methods have been proposed, including optimized training strategies \citep{Qu2021, Ren2021}, enhanced similarity measurements \citep{Ren2021a}, and improved training efficiency \citep{Hofstaetter2021}, none have investigated the impact of associating multiple positive passages with each question during training.

In this paper, we focus on pairing multiple positive passages with each question when training the dual BERT encoder. The intuition is straightforward: we hypothesize that the dominant number of negative passages during training (e.g., 1 positive vs. 255 negatives per question in DPR) may erode the model's ability to identify relevant passages at inference time. By exposing the model to more positive passages, we reformulate training as a binary classification task, where the model learns to distinguish between positive and negative passages under a smaller positive-negative imbalance. Experimental results on several QA datasets consistently show that our method improves retrieval accuracy while significantly reducing the required batch size, enabling the model to be trained on a single GPU.

\section{Methodology}

This section presents the method that incorporates multiple positive passages with each question to train the DPR model. As a preliminary step, we first introduce some background of the DPR model.

\subsection{The DPR model}

DPR uses two separate encoders, \( E_P(\cdot) \) and \( E_Q(\cdot) \), to map text passages and questions to a shared vector space. Both encoders are based on the BERT (base, uncased) model, and the representation at the \( [CLS] \) token is used as the output. 

The training objective is to ensure that the distance between relevant question-passage pairs is smaller than the irrelevant ones. This distance (or similarity) between a question and a passage is measured by the inner product of their vector representations:

\begin{equation}
	\text{sim}(q, p) = E_Q(q)^\top E_P(p)
\end{equation}

Suppose there is one positive (relevant) passage \( p_i^+ \) and \( n \) negative (irrelevant) passages \( p_{i,j}^- \), where \( j \in \{1, 2, \ldots, n\} \), for each question \( q_i \). DPR is then optimized using the negative log likelihood (NLL) of the positive passage:

\begin{equation}\label{eq:old-training}
	\begin{split}
		&\mathcal{L}(q_i, p_i^+, p_{i,1}^-, \dots, p_{i,n}^-) = \\
		&- \log \frac{e^{\text{sim}(q_i, p_i^+)}}{e^{\text{sim}(q_i, p_i^+)} + \sum_{j=1}^{n} e^{\text{sim}(q_i, p_{i,j}^-)}}.
	\end{split}
\end{equation}

During the actual training process, each question is paired with one positive passage and one hard negative passage. By taking advantage of the in-batch negatives trick \citep{Yih2011}, all positive and negative passages associated with other questions are treated as negative passages for the current question, enabling efficient computation and improved performance. Assuming the batch size is \(B\), for each question-passage training sample, the ratio of positive passages to negative passages is roughly \(1 / 2B\). In the next subsection, we discuss the differences that after pairing each question with multiple positive passages.

\subsection{Pairing each question with more positive passages}

Instead of a single positive passage, we pair each question with \(m > 1\) positive passages to train the model. After applying in-batch training, there are \( (m + 1) \times (B - 1) + 1 \) negative passages for each question, and the ratio of positive to negative passages is approximately \(1 / B\), doubling the proportion compared to the original DPR model. We hypothesize that this provides the model with more positive feedback during training, which could be beneficial for improving performance.

Since more positive passages are introduced, we treat the training process as a binary classification task, where the model is expected to judge each passage as either positive or negative with respect to a question. To optimize the model, we discard the NLL loss formulated in Eq.~\eqref{eq:old-training} and instead use binary cross-entropy (BCE) loss:

\begin{equation}\label{eq:training}
	\begin{split}
		&\mathcal{L}(q_i, p_{i,1}^+, \dots, p_{i,m}^+, p_{i,1}^-, \dots, p_{i,n}^-) = \\
		& - \sum_{k=1}^{m} \log \sigma\left( \text{score}(q_i, p_{i,k}^+) \right) \\
		& - \sum_{j=1}^{n} \log \left( 1 - \sigma\left( \text{score}(q_i, p_{i,j}^-) \right) \right)
	\end{split}
\end{equation}

In Eq.~\eqref{eq:training}, \( \text{score}(q, p) \) refers to the softmax-scaled inner product similarity \( \text{sim}(q, p) \), and \( \sigma(\cdot) \) is the sigmoid function. These configurations are used to stabilize the training process.

\section{Experimental setup}

This describes the data used in our experiments and the training configurations.

\subsection{Data preparing}

The training datasets and source documents are the same as those used in \citep{Karpukhin2020}. The source documents are constructed using Wikipedia English articles (Dec 20, 2018 dump). These documents consist of 21,015,324 passages, with each passage containing 100 words. Details of the training datasets are provided below.

\textbf{CuratedTREC (TREC)} \citep{Baudis2015} is an improved QA training and benchmark dataset derived from the TREC QA tracks. Some answers are expressed using regular expression patterns.

\textbf{WebQuestions (WebQ)} \citep{Berant2013} was crafted using Google Suggest API, and all questions begin with a wh-word.

\textbf{SQuAD 1.1} \citep{Rajpurkar2016} contains 107,785 question-answer pairs derived from 536 Wikipedia articles via crowdsourcing.

\textbf{TriviaQA} \citep{Joshi2017} is a reading comprehension dataset consisting of 95,000 trivia questions. Each question is associated with six evidence documents on average.

\textbf{Natural Question (NQ)} \citep{Kwiatkowski2019} is a real-world question answering benchmark dataset with questions mined from Google search queries and answers annotated from Wikipedia articles.

We retain the passage selection strategy as demonstrated in \citep{Karpukhin2020}, and discard samples that could cause the training process failure. Table \ref{tab:dataset_split} lists the actual number of questions in each dataset for training the model.

\begin{table}[h]
	\centering
	\small
	\begin{tabular}{l c}
		\hline
		\textbf{Dataset} & \textbf{questions used for training} \\
		\hline
		TREC             & 1,117         \\    
		WebQ             & 2,448          \\ 
		SQuAD            & 70,096          \\  
		TriviaQA         & 60,368           \\  
		NQ               & 58,880            \\
		\hline
	\end{tabular}
	\caption{Questions used for training the model in each dataset.}
	\label{tab:dataset_split}
\end{table}

\begin{table*}[!t]
	\small
	\centering
	\setlength{\tabcolsep}{5pt} 
	\begin{tabular}{ll|ccccc|ccccc}
		\hline
		\multirow{2}{*}{\textbf{Training}} &\multirow{2}{*}{\textbf{Retriever}} & \multicolumn{5}{c|}{\textbf{Top-20}} & \multicolumn{5}{c}{\textbf{Top-100}} \\
		\cline{3-12}
		& & NQ & TriviaQA & WQ & TREC & SQuAD & NQ & TriviaQA & WQ & TREC & SQuAD \\
		\hline
		None & BM25 & 59.1 & 66.9 & 55.0 & 70.9 & 68.8 & 73.7 & 76.7 & 71.1 & 84.1 & 80.0 \\
		\hline
		\multirow{3}{*}{Single} 
		& DPR & 78.4 & 79.4 & 73.2 & 79.8 & 63.2 & 85.4 & 85.0 & 81.4 & 89.1 & 77.2 \\
		& BM25 + DPR & 76.6 & 79.8 & 71.0 & 85.2 & \textbf{71.5} & 83.8 & 84.5 & 80.5 & 92.7 & \textbf{81.3} \\
		& $\text{DPR}^+$ & \textbf{80.4} & \textbf{79.9} & \textbf{76.5} & 83.6 & 52.7 & \textbf{86.8} & \textbf{85.6} & \textbf{83.4} & 92.7 & 69.2 \\
		\hline
		\multirow{2}{*}{Multi} 
		& DPR & 79.4 & 78.8 & 75.0 & \textbf{89.1} & 51.6 & 86.0 & 84.7 & 82.9 & 93.9 & 67.6 \\
		& BM25 + DPR & 78.0 & \textbf{79.9} & 74.7 & 88.5 & 66.2 & 83.9 & 84.4 & 82.3 & \textbf{94.1} & 78.6 \\
		\hline
	\end{tabular}
	\caption{Top-20 and Top-100 retrieval accuracy across test datasets. The accuracy is calculated as the percentage of top \( 20 / 100 \) retrieved passages that contain the answer. \textit{Single} and \textit{Multi} denote that the retriever was trained using one or combined datasets (all excluding SQuAD). Bold numbers indicate the best performance.}
	\label{tab:retrieval-results}
\end{table*}

\subsection{Training setup}

We pair each question with up to 3 positive passages and 1 hard negative passage, and use the in-batch negative trick to train the model. Since some questions have fewer than 3 associated positive passages, we dynamically assign the maximum available number of positives for those cases. The batch size is set to 16, with 100 training epochs for TREC and WebQ, and 40 for SQuAD, TriviaQA, and NQ. We use the Adam optimizer with a learning rate of \( 10^{-5} \), linear scheduling with warm-up, and a dropout rate of 0.1. All experiments were conducted on a single NVIDIA RTX 5080 GPU with 16 GB of VRAM. 

\section{Experiment results}

This section demonstrates the evaluation results of the proposed model along with analysis of its effectiveness.

\subsection{Main results}

The top \( k \) (\( k \in \{20, 100\} \)) retrieval accuracy of different models are shown in Table \ref{tab:retrieval-results}. \( \text{DPR}^+ \) denotes our proposed model, while the others are from \citep{Karpukhin2020}. Single and Multi indicate whether the model was trained on individual or combined datasets (all but except SQuAD). BM25 + DPR is a linear combination of the BM25 and DPR models, as described in \citep{Karpukhin2020}.

The results show that \( \text{DPR}^+ \) achieves the best performance on the NQ, TriviaQA, and WebQ datasets, even without multi-dataset training or BM25 model combination. Notably, our model outperforms the Single DPR baseline on all datasets except SQuAD, clearly confirming its effectiveness. Additionally, thanks to a significantly smaller batch size, our model can be trained on a single GPU with 16 GB of VRAM, whereas training the original DPR model with a batch size of 128 requires \( 8 \times 32 \) GB GPUs. This highlights the improved efficiency of \( \text{DPR}^+ \).

We suspect that the low performance of \( \text{DPR}^+ \) on the SQuAD dataset is due to an inadequate number of positive passages. Based on the number of positive passages associated with each question, we classified the questions into three groups, as shown in Table~\ref{tab:positive}. In the table, \( p_1 \) denotes the number of questions paired with only one positive passage; the same terminology applies to \( p_2 \) and \( p_3 \). The symbol \( \delta \) represents the proportion of \( p_3 \) in the total. As illustrated in the table, only 49.1\% of the questions in the SQuAD dataset are paired with three positive passages. This data deficiency may hinder \( \text{DPR}^+ \) from fully exploiting the benefits of multiple positive passages during training.

\begin{table}[!t]
	\centering
	\resizebox{\columnwidth}{!}{%
		\begin{tabular}{l c c c c c}
			\hline
			\multirow{2}{*}{\textbf{Dataset}} & \multirow{2}{*}{\textbf{$\delta$}} & \multicolumn{4}{c}{\textbf{question count}} \\
			                                     \cline{3-6}
			&                                    &$p_1$ &$p_2$ &$p_3$ &total     \\
			\hline
			TREC             & 82.1\%            & 89   & 111  & 920  & 1,120      \\
			WebQ             & 74.1\%            & 365  & 273  & 1,826 & 2,464      \\
			SQUAD            & 49.1\%            & 14,842 & 20,834 & 34,420 & 70,096 \\
			TriviaQA         & 79.5\%            & 6,336 & 6,029 & 48,035 & 60,400    \\
			NQ               & 67.7\%            & 9,577 & 9,455 & 39,848 & 58,880     \\
			\hline
		\end{tabular}
	}
	\caption{Question counts with respect to the number of positive passages associated with each question in the training datasets. \( p_1 \), \( p_2 \), and \( p_3 \) denote the number of questions paired with 1, 2, and 3 positive passages, respectively. \( \delta \) represents the proportion of \( p_3 \) in the total number of questions.}
	\label{tab:positive}
\end{table}

\subsection{Ablation study}

We selected the TREC dataset and varied the maximum number of positive passages per question to 1, 2, and 3, respectively, to examine the impact of positive passages on performance. We also included the results of DPR-Single \citep{Karpukhin2020} trained with a batch size of 16 to assess the influence of using the BCE training loss. For simplicity, we reused the same encoded source files from our previous experiments. The final results are presented in Table~\ref{tab:exp-abl}.

\begin{table}[!t]
	\centering
	\resizebox{\columnwidth}{!}{
		\begin{tabular}{l c c}
			\hline
			\textbf{Model} & \textbf{Top-20 accuracy} & \textbf{Top-100 accuracy} \\
			\hline
			DPR-Single & 80.8 & 89 \\
			$\text{DPR}^+_1$   & 80.4 &   89.6         \\
			$\text{DPR}^+_2$   & 83.9 &  91.6           \\
			$\text{DPR}^+_3$   & 83.6 & 92.7             \\ \hline
		\end{tabular}
	}
	\caption{Top-20 and Top-100 retrieval accuracy for different methods. \( \text{DPR}^+_i \) denotes the \( \text{DPR}^+ \) model trained with \( i \) positive passages paired with each question. The results for DPR-Single are reported using a batch size of 16 during training.}
	\label{tab:exp-abl}
\end{table}

In Table~\ref{tab:exp-abl}, \( \text{DPR}^+_1 \), \( \text{DPR}^+_2 \), and \( \text{DPR}^+_3 \) represent \( \text{DPR}^+ \) trained with 1, 2, and 3 positive passages associated with each question, respectively. We observe that performance improves as more positive passages are incorporated, and the difference between DPR-Single and \( \text{DPR}^+_1 \) is relatively small. The improvement becomes more pronounced in terms of top-100 accuracy. These results suggest that the BCE loss has only a subtle impact on performance, whereas pairing each question with more positive passages leads to clear performance gains.

\section{Conclusion}

In this paper, we present a simple yet effective strategy that pairs multiple positive passages with each question to enhance the DPR model. By formulating training as a binary classification task, where each passage is judged as positive or negative, the model is optimized using the BCE loss. Empirical results demonstrate that the proposed method consistently improves retrieval accuracy while significantly reducing the hardware requirements for training.

\section*{Limitations}

Due to hardware constraints, we were only able to train the model with up to three positive passages per question and a batch size of 16. The effects of using more positive passages or larger batch sizes remain unexplored. Further investigation is required to understand the trade-off between batch size and the number of positive passages for optimal performance.

Moreover, while our method is simple and easy to implement, it may have limitations in further improving retrieval accuracy compared to more sophisticated approaches.

\section*{Acknowledgments}

The author would like to thank anyone who offered constructive suggestions.

\bibliography{custom.bib}

\end{document}